\documentclass[12pt,a4paper]{article}
\usepackage[vmargin=20mm,hmargin=20mm]{geometry}

\def\Cpp{{C\nolinebreak[4]\hspace{-.05em}\raisebox{.2ex}{+\hspace{-.2em}+}}} 
\usepackage{todonotes}
\definecolor{crs4blue}{RGB}{12,77,162}
\usepackage{mwe}
\usepackage[T1]{fontenc}
\usepackage[utf8]{inputenc}
\usepackage{subcaption}
\usepackage{amsmath,amssymb,amsfonts}
\usepackage{url}
\usepackage{authblk}

\usepackage[sc]{mathpazo}

\usepackage{booktabs}
\usepackage{dcolumn}
\newcolumntype{d}[1]{D{.}{.}{#1}}
\usepackage{multirow}

\usepackage{paralist}

\usepackage{float}
\usepackage{listings}
\floatstyle{ruled}
\newfloat{lstfloat}{htbp}{lop}
\floatname{lstfloat}{Listing}
\begin{document}

\title{Hiding Latencies in Network-Based Image Loading for Deep
  Learning}

\author{Francesco Versaci}
\author{Giovanni Busonera}
\affil{CRS4, Cagliari, Italy}

\maketitle

\begin{abstract}
  In the last decades, the computational power of GPUs has grown
  exponentially, allowing current deep learning (DL) applications to
  handle increasingly large amounts of data at a progressively higher
  throughput.
  However, network and storage latencies cannot decrease at a similar
  pace due to physical constraints, leading to data stalls, and
  creating a bottleneck for DL tasks.
  Additionally, managing vast quantities of data and their associated
  metadata has proven challenging, hampering and slowing the
  productivity of data scientists.
  Moreover, existing data loaders have limited network support,
  necessitating, for maximum performance, that data be stored on local
  filesystems close to the GPUs, overloading the storage of computing
  nodes.

  In this paper we propose a strategy, aimed at DL image applications,
  to address these challenges by: storing data and metadata in fast,
  scalable NoSQL databases; connecting the databases to
  state-of-the-art loaders for DL frameworks; enabling high-throughput
  data loading over high-latency networks through our out-of-order,
  incremental prefetching techniques. To evaluate our approach, we
  showcase our implementation and assess its data loading capabilities
  through local, medium and high-latency (intercontinental)
  experiments.%
\end{abstract}


\section{Introduction}
Over the last two decades, the rapid increase in GPU computational
power has transformed the field of machine learning. This surge in
processing capabilities has allowed deep learning (DL) models to
manage vast datasets and conduct intricate computations with unmatched
efficiency. Consequently, DL techniques have gained widespread
traction across various sectors, leading to advancements in areas such
as cybersecurity, natural language processing, bioinformatics,
robotics and control, and medical information processing, among
others~\cite{dl-review}.

However, despite these advancements in computational power, the
performance of DL systems is increasingly constrained by bottlenecks
related to data movement and access. While GPUs continue to achieve
remarkable gains in processing throughput, which is matched by an
increase in bandwidth for networking and storage, the latencies in
these areas cannot decrease at the same pace due to inherent physical
limitations. These discrepancies result in significant data stalls, as
the transfer of data between storage systems, memory, and processing
units becomes a critical bottleneck, leading to inefficient resource
utilization in DL workflows~\cite{data-stalls-google, data-stalls-ms}.

Furthermore, managing the vast amounts of data and associated metadata
required for DL has become an increasingly complex challenge,
significantly impacting the productivity of data scientists. As
datasets scale from sources such as ImageNet~\cite{imagenet}, which
comprises millions of images and spans hundreds of gigabytes, to more
extensive datasets like LAION-5B~\cite{laion}, containing billions of
images and reaching hundreds of terabytes, the exponential growth in
data volume places substantial strain on traditional filesystem-based
approaches. These systems are often inflexible and ill-suited for
dynamic data management requirements. The limitations of these
approaches are particularly evident when tasks such as balancing class
distributions~\cite{imbalance}, or adjusting the proportions of
training, validation, and test sets, must be performed, especially
when considering metadata to avoid introducing unwanted biases. The
static nature of traditional filesystems hampers the ability to
efficiently update or modify datasets, restricting the flexibility
needed for iterative development and fine-tuning of DL models.

Moreover, DL applications often involve datasets comprising numerous
small inputs that are fully scanned and randomly permuted at each
training epoch. This access pattern diminishes the benefits of caching
since the data is continuously reloaded and
reshuffled~\cite{data-locality}. Additionally, high-performance
computing (HPC) storage systems are typically optimized for sequential
access to large files, which contrasts sharply with the small, random
access patterns typical of DL workloads. As a result, parallel file
systems such as GPFS or Lustre in HPC environments struggle to handle
these workloads efficiently~\cite{hpc-sota}.
A common mitigation strategy is to maintain local copies of datasets
on all compute nodes. However, this approach introduces substantial
storage overhead and capacity constraints. An alternative is to
partition the dataset into shards distributed across nodes, which
becomes essential when the dataset exceeds the storage capacity of
individual nodes. Yet, ensuring unbiased data sharding poses
significant challenges~\cite{hpc-sota}. This strategy also compromises
the ability to perform uniform random shuffling, which can negatively
impact training performance and model accuracy~\cite{shuffling}.

To address these limitations, specialized data loading systems and
strategies have been developed, as elaborated in the next
section. However, these methods have critical shortcomings, leaving
unresolved issues such as:
\begin{itemize}
\item Decoupling of data and metadata, leading to potential
  inconsistencies;
\item Inflexibility of record file formats, which impose constraints
  on shuffling;
\item Limited support for network-based data loading, resulting in
  local storage overload on computing nodes.
\end{itemize}
In response to these issues \cite{cass-eddl} proposed leveraging
scalable NoSQL databases to store both data and metadata, with
preliminary performance evaluations conducted within the DeepHealth
Toolkit~\cite{deephealth}, a DL framework tailored for biomedical
applications.
In this work, we build upon this concept by introducing and evaluating
a novel data loader. Specifically, our key contributions are as
follows:
\begin{itemize}
\item We develop an efficient data loader implemented in \Cpp\ with a
  Python API, designed to integrate seamlessly with
  Cassandra-compatible NoSQL databases and NVIDIA DALI~\cite{dali,
    torch-ffcv-dali}. This loader supports data loading across the
  network and is compatible with popular DL frameworks such as PyTorch
  and TensorFlow.
\item We introduce out-of-order, incremental prefetching techniques that
  enable high-throughput data loading, even in high-latency network
  environments;
\item We conduct a comprehensive evaluation of our approach,
  demonstrating its implementation and benchmarking its performance
  through extensive experiments in local, medium and high-latency
  settings, comparing it against the state-of-the-art tools.
\end{itemize}
Note that, as DALI is primarily optimized for image processing, our
examples will focus on DL applications involving images; however, the
techniques presented are of general applicability.

\section{Background and related work}
In this section, we begin with an overview of record file formats in
DL. Next, we review state-of-the-art data loading software,
highlighting their advantages and limitations. Finally, we introduce
the NoSQL databases that will be utilized by our data loader.

\subsection{Record file formats}
Many DL applications, such as image classification, exhibit limited
temporal and spatial locality due to their scan-and-reshuffle data
access patterns~\cite{data-locality}. This lack of locality impedes
optimization strategies such as block reading, which involves
retrieving multiple images in a single request. When adopted, block
reading can help mask latency, alleviate stress on the filesystem, and
ultimately improve throughput in both storage and network systems.
To leverage this optimization, despite the lack of inherent spatial
locality, engineers have developed file formats that artificially
enforce spatial locality by grouping (unrelated) files together in
record files. Examples include TFRecord~\cite{tfdata} (initially
developed for TensorFlow and also supported by
PyTorch\footnote{\url{https://pytorch.org/data/main/generated/torchdata.datapipes.iter.TFRecordLoader.html}}),
RecordIO (designed for the now archived MXNet
framework\footnote{\url{https://attic.apache.org/projects/mxnet.html}}),
Beton (developed within the FFCV project~\cite{ffcv}) and MDS
(developed within Databricks's MosaicML platform,
see~\S\ref{sec:streamingdataset}), the first two being also supported
by NVIDIA DALI.

However, optimization algorithms like Stochastic Gradient Descent and
Adam require uniform shuffling of data for optimal
convergence~\cite{shuffling}. Block reading conflicts with this
requirement, leading to the implementation of workarounds in data
loaders. For instance, they may load a window of data several times
larger than the desired minibatch size into memory and shuffle
internally, ensuring that minibatches vary across epochs. Even so, the
resulting shuffle is not uniform, as images stored close together in a
record file will always appear in nearby minibatches.

Nevertheless, the primary drawback of the file-batching approach is
that it further rigidifies the dataset. Writing record files is
time-consuming, consumes additional storage space, and makes it even
more challenging to modify datasets -- an already cumbersome task when
dealing with numerous files in a filesystem.

\subsection{State-of-the-art data loaders}
\subsubsection{NVIDIA DALI}
\label{sec:nvidia-dali}
A major inefficiency in a typical naive image classification
workflow\footnote{E.g.,
  \url{https://pytorch.org/tutorials/beginner/transfer_learning_tutorial.html}}
stems from the fact that image loading and decoding are managed by
high-level routines written in Python. Due to Python’s Global
Interpreter Lock (GIL), which limits parallel multi-threading, data
loading suffers from considerable serialization and copying overhead
between processes~\cite{parallel-python, pytorch-rfc}.
Additionally, pre-processing operations such as resizing, rotating,
cropping, and normalization are typically performed on the CPU, which
can become a bottleneck, causing delays as the faster GPUs remain idle
while waiting for data.

NVIDIA DALI is a state-of-the-art data loader~\cite{dali,
  torch-ffcv-dali}, which addresses these inefficiencies by managing
the entire pipeline of loading, decoding, and pre-processing
images. It leverages GPU acceleration for decoding and pre-processing
tasks, significantly reducing CPU bottlenecks. DALI integrates
seamlessly with both PyTorch and TensorFlow, offering a versatile
solution for DL workflows. Its asynchronous, pipelined execution model
with prefetching ensures efficient data handling and minimizes idle
GPU time. As free software written in \Cpp{}, with Python bindings,
DALI is modular and extensible, allowing users to customize it with
their own plugins for specialized tasks.

One current limitation of DALI is its limited support for data loading
over networks. As of this writing, it has only recently introduced
experimental support for data loading from S3, and its performance
remains
suboptimal\footnote{\url{https://github.com/NVIDIA/DALI/issues/5551}}.
Additionally, DALI does not currently support reading data from more
structured sources, such as databases.

\subsubsection{TensorFlow's tf.data}
The \texttt{tf.data} module \cite{tfdata} is an efficient framework
for data loading and processing, developed as part of the TensorFlow
ecosystem. Notably, it is exclusive to TensorFlow and is not supported
by PyTorch. Like DALI, it is implemented in \Cpp{} for performance
reasons but provides a Python API, allowing seamless integration with
TensorFlow workflows. One of the key strengths of \texttt{tf.data} is
its ability to efficiently manage parallel data loading and
preprocessing, significantly reducing input pipeline bottlenecks
during training on large datasets. The module supports various data
sources, including local file systems and networked storage,
facilitating distributed data handling and enabling the construction
of scalable and flexible data input pipelines. It offers a range of
optimizations, such as prefetching, caching, sharding, and use of
record files (TFRecord) to improve overall throughput in DL workflows.

However, its native support for network-based data loading is limited
to Google Cloud services, which may pose constraints for users relying
on alternative cloud platforms or custom data storage solutions.

\subsubsection{tf.data service}
The \texttt{tf.data service} \cite{tf-data-service}, still experimental,
offers an alternative approach for data loading in TensorFlow. This
service extends the \texttt{tf.data} API by decoupling data loading
and preprocessing from model training, thus enabling efficient scaling
of data input pipelines across multiple workers. By offloading data
preprocessing and distribution to a centralized service, hosted on
dedicated nodes, the \texttt{tf.data service} aims at enhancing
parallel data loading and transformation, helping to alleviate input
pipeline bottlenecks.
It is worth noting that this service can also be utilized to enable
network-based data loading, with the dispatcher and workers supplying
data to the nodes responsible for model training.

\subsubsection{MosaicML Streaming Dataset}
\label{sec:streamingdataset}
MosaicML Streaming Dataset (SD) library~\cite{mosaicml2022streaming}
is a Python package designed for efficient and scalable handling of
large datasets, especially in machine learning and data processing
workflows. It supports datasets that exceed the capacity of a single
node by using streaming techniques to load and locally buffer data
dynamically during training or evaluation. Key features include the
ability to mix multiple data sources with configurable weighting,
flexible handling of diverse data types (e.g., images, text, video,
and multimodal data), and true determinism to facilitate debugging and
reproducibility.

Note that to utilize SD, data must be converted into a proprietary
record format (known as Mosaic Data Shard). This format organizes data
samples into large files called shards, accompanied by an index file
containing metadata. The metadata enable efficient partitioning of
samples across nodes and GPUs prior to training, reducing redundant
downloads and enhancing performance.

\subsubsection{Deep Lake}
Deep Lake~\cite{deeplake} is a specialized database optimized for DL
applications, combining the features of data lakes and vector
databases to efficiently manage a wide range of data types, including
text, audio, video, images, PDFs, embeddings, and annotations. It is
particularly suited for the development of Large Language Model
applications and the training of DL models. Deep Lake utilizes a
proprietary record file format, Tensor Storage Format, and supports
remote data access through AWS S3, Google Cloud Storage, Azure, or its
own Activeloop Storage. The platform offers serverless, scalable
storage solutions, enabling users to manage diverse datasets in a
unified environment.

A key limitation of Deep Lake is that its high-performance \Cpp{} data
loader for distributed training at scale is closed-source and only
available in the premium enterprise version, which requires a paid
subscription. Additionally, this faster data loader does not support
self-hosting, forcing users to rely on external cloud storage
services. This limitation may pose challenges for organizations with
specific data residency requirements or those seeking to minimize data
transfers outside their internal networks.

\subsubsection{MADlib}
MADlib~\cite{madlib} is an open-source, in-database machine learning
library originally developed for scalable and parallelized machine
learning on relational databases. Its core concept is to bring machine
learning closer to where the data resides, enabling analytics and
training operations to be performed directly inside the database. This
approach minimizes the need for large-scale data transfers and
leverages the parallel processing capabilities of modern relational
database management systems (RDBMS). Initially designed for
traditional machine learning algorithms like logistic regression,
decision trees, and k-means clustering, MADlib has evolved to support
more advanced analytics.  Recent developments have integrated support
for DL frameworks like Keras, allowing the use of neural networks
alongside conventional ML algorithms. However, DL support in MADlib is
still at an early stage and has several limitations:
\begin{itemize}
\item Images are stored as uncompressed tensors;
\item Only image classification is currently supported (no
  segmentation);
\item Minibatches are solidified, i.e., they are saved together as a
  row in the DB.
\end{itemize}

Interestingly, MADlib's trajectory can be interpreted as an inverse
approach compared to the methodology examined in our work. While
MADlib started as an RDBMS-centric solution for machine learning and
is gradually expanding its capabilities toward DL, this work connects
traditional DL data loaders to NoSQL database for image storage and
retrieval over the network. This highlights a convergence where both
approaches -- extending database-centric tools to support DL, and
using specialized databases to manage the growing complexity of DL
data -- meet similar challenges from opposite directions. The shared
goal is optimizing data management and processing efficiency in
large-scale, data-intensive learning scenarios.

\subsection{Cassandra-compatible databases}
Our data loader, outlined in the following section, retrieves data
from Cassandra-compatible databases, which are summarized below.

\subsubsection{Apache Cassandra}
Apache Cassandra is a secure, distributed, and decentralized NoSQL
database system known for its high scalability and fault tolerance. It
is specifically designed to support geographically distributed
deployments across multiple data centers, ensuring high availability
and resilience. Written in Java and open-source, Cassandra is
optimized for low-latency operations, typically delivering response
times in the range of single to low double-digit milliseconds for
small data transactions. These characteristics have made it a popular
choice in the big data industry, with notable adopters including
Netflix and Spotify.

\subsubsection{ScyllaDB}
ScyllaDB is a real-time, big-data NoSQL database engineered to be
API-compatible with Cassandra. Developed as open-source software in
\Cpp, it offers significant performance enhancements over Cassandra,
particularly in terms of reduced latency and lower variability. This
improvement is largely due to the absence of a Java garbage collector
and the use of a shard-per-core architecture. These optimizations make
ScyllaDB well-suited for high-performance workloads, as also proven by
its adoption by major companies such as Discord, Ticketmaster, and
Rakuten.

\section{High-performance network data loading: design principles and
  implementation}
The design of our data loader was guided by the following critical
objectives, each aimed at addressing the unique challenges posed by
modern deep learning workflows:

\begin{itemize}
\item \textbf{Integration of data and metadata storage:} By tightly
  coupling data with its associated metadata, we minimize the risk of
  inconsistencies arising from discrepancies between the two. This
  design choice ensures that datasets remain reliable and
  self-contained, eliminating potential errors during dataset update.
\item \textbf{Provision of fast, flexible, and scalable network
    access:} Our architecture supports high-throughput access to data,
  ensuring that input pipelines do not become a bottleneck in
  training. It is designed to scale seamlessly across a range of
  network configurations, from local clusters to cloud-based
  environments, while maintaining low latency and robust performance.
\item \textbf{Ensuring unrestricted random access:} The ability to
  access any data sample at random is crucial for shuffling datasets
  uniformly during training. Our loader is built to guarantee this
  capability, even when handling large-scale datasets stored across
  distributed systems, ensuring that training remains unbiased and
  effective.
\item \textbf{Simplifying data partitioning and data distribution in
    parallel settings:} By separating storage management from dataset
  partitioning, our design offers flexibility for dynamic splitting
  strategies like cross-validation while maintaining a consistent
  retrieval interface. It also streamlines data distribution across
  nodes or GPUs, reducing overhead and ensuring efficient operation in
  distributed training setups.
\end{itemize}

Together, these design principles underpin a robust, efficient, and
user-friendly data loading solution tailored for the demands of DL
applications, particularly those involving large-scale image
classification tasks.
To implement the design principles outlined above, we developed our
data loader as a plugin that bridges NVIDIA DALI and a
Cassandra-compatible NoSQL database. This approach enables
high-performance data retrieval, ensuring scalability and reliability
for modern DL workloads. By implementing the data loader in \Cpp{}, we
eliminate the interprocess communication overhead typically associated
with Python-based loaders, providing significant improvements in data
throughput and latency.

\subsection{Data flow and model}
\begin{figure}
  \centering
  \includegraphics[width=.9\linewidth]{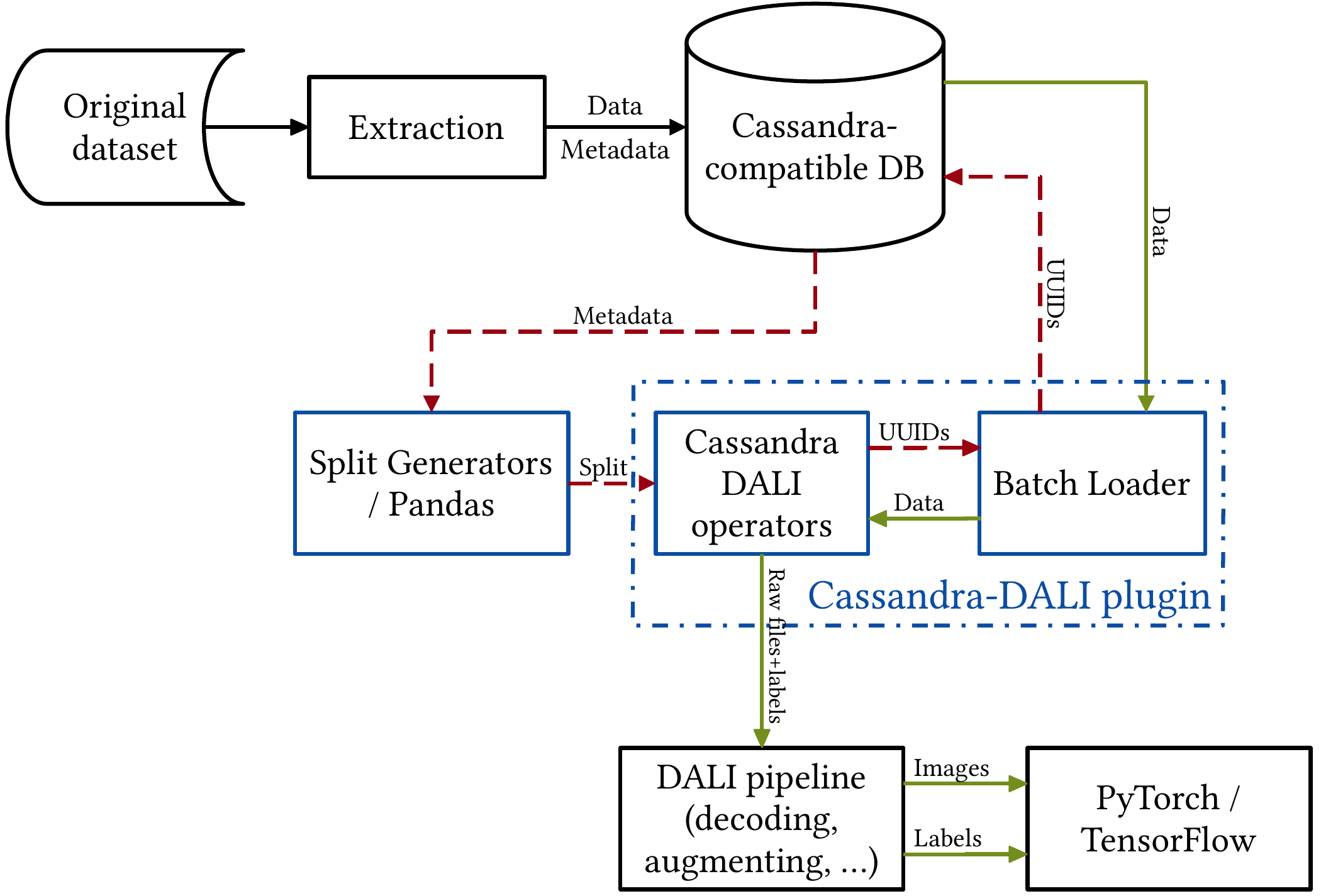}
  \caption{High-level architecture of our data loader.}
  \label{fig:architecture}
\end{figure}

The data flow of DL applications utilizing our data loader is depicted
in Fig.~\ref{fig:architecture} and proceeds as follows:
\begin{itemize}
\item Data are extracted from the original dataset and stored, along
  with associated metadata, in a Cassandra-compatible database;
\item Each image is uniquely identified using a Universally Unique
  Identifier (UUID);
\item Splits, represented as lists of UUIDs, can be automatically
  generated based on target values and constraints defined by the
  metadata;
\item When needed, data are efficiently retrieved using their UUIDs,
  passed to the DALI pipeline for preprocessing, and then fed into the
  DL engine (either PyTorch or TensorFlow).
\end{itemize}

This architecture enables data loading over the network using TCP,
allowing full random access to datasets. Additionally, by leveraging
Cassandra or ScyllaDB for storage, it offers significant advantages
such as easy scalability and secure data access through SSL. It also
allows for the definition of roles with detailed permissions and
supports straightforward geographic replication.

To optimize performance and simplify queries, we store data and
metadata in separate tables. This separation improves retrieval
efficiency: metadata, which is solely used for creating data splits,
can be queried independently of the data accessed during the training
process. However, in order to ensure database consistency, data and
metadata associated with the same image are inserted atomically using
Cassandra’s BatchStatement.

As an example of image classification that involves a complex set of
metadata, we present SQL tables in Listing~\ref{lst:data-model} that
are applicable to medical imaging, specifically for tumor detection in
digital pathology~\cite{dl-tum}. In this context, gigapixel images are
divided into small patches, with each patch being labeled to indicate
the degree or severity of cancerous tissue (Gleason score).

It is important to note that our data loader is designed with
flexibility beyond standard image classification tasks. It accepts
features stored as a generic BLOB, allowing for any file format
decodable by DALI (e.g., JPEG, JPEG 2000, TIFF, PNG, etc.), which also
offers the option to include custom decoders if needed. Annotations
are optional and can be provided as integer values (for
classification) or as another BLOB, enabling support for tasks like
multilabel classification (where labels can be stored as serialized
NumPy tensors) and semantic segmentation (where masks are stored as
images, for instance, in PNG format).

\begin{lstfloat}
\caption{Example of SQL data model for tumor detection}
\label{lst:data-model}
\begin{lstlisting}
CREATE TABLE patches.£metadata£(
  patient_id text,
  slide_num int,  // patients can have
                  // several slides
  x int,  // coordinates
  y int,  // within the slide
  label int,  // Gleason score
  patch_id uuid,
  PRIMARY KEY ((patch_id))
);

CREATE TABLE patches.£data£(
  patch_id uuid,
  label int,  // Gleason score
  data blob,  // image/tensor file
              // (JPEG, TIFF, NPY, etc.)
  PRIMARY KEY ((patch_id))
);
\end{lstlisting}
\end{lstfloat}

\subsection{Automatic Split Creation}
Creating dataset splits, such as training, validation, and test sets,
requires careful consideration of metadata to ensure both data
independence and balance. A key consideration in this process is
maintaining entity independence, where specific entities -- such as
individuals, groups, or sessions -- must be assigned to distinct
splits to prevent data leakage and preserve the integrity of model
evaluation. For instance, in medical datasets, it is crucial to assign
patients to separate splits to avoid bias and ensure generalizability.
Additionally, it is common practice to adjust class distributions
within the splits to reflect desired proportions of each class, which
may involve modifying class weights to address imbalances. However,
managing these tasks with standard workflows, i.e., including the
reorganization of subdirectories or the recreation of TFRecord files,
can be highly labor-intensive and prone to error, underscoring the
need for automated solutions to streamline the process.

Our plugin automates both the creation of dataset splits and the
corresponding data loading process, decoupling storage from splits and
eliminating the need for manual creation of TFRecord files or
subdirectory reorganization. By automating both split creation and
data loading, it enhances workflow efficiency, enabling users to focus
on model development without the complexities of managing dataset
splits or file organization. This automation not only simplifies the
integration process but also improves reproducibility, facilitating
consistent and efficient handling of complex datasets.

\subsection{Multi-threaded asynchronous data loading}
To optimize data retrieval efficiency, we leverage extensive
parallelization. Images are retrieved asynchronously across multiple
threads and TCP connections, thereby minimizing overall latency. Each
TCP connection can handle up to 1024 concurrent requests, with the
number of connections being a tunable parameter. Once the images are
retrieved, batches of data are assembled in shared memory, which
eliminates the need for additional copying and accelerates the
process.

\begin{figure*}
  \centering
  \includegraphics[width=\linewidth]{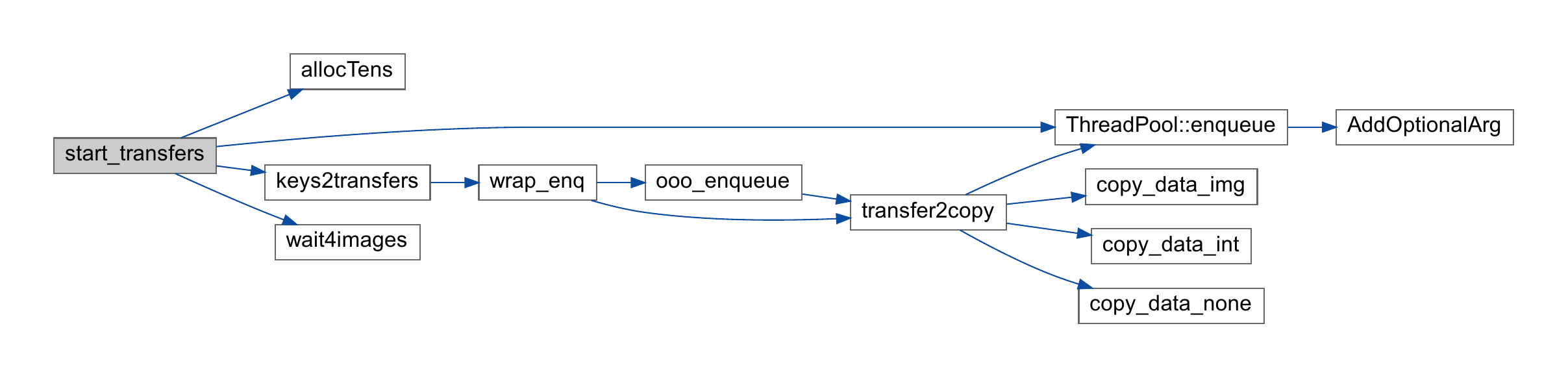}
  \caption{Call graph of our data loader.}
  \label{fig:call-graph}
\end{figure*}

In detail, our batch loading workflow starts with the batch loader
receiving a list of UUIDs, which it then uses to send all requests to
the Cassandra driver at once. Communications for different batches are
handled concurrently via a thread pool. To manage these requests
efficiently, multiple low-level I/O threads are employed, each
utilizing two TCP connections. Results are processed through
callbacks, which minimizes latency by eliminating busy waiting. After
all results for a batch have been received, the output tensor is
allocated contiguously in a single operation. Data is then copied into
the output tensor concurrently, again using a thread pool. The batch
becomes available for output as soon as the copying is complete. The
call graph illustrating this workflow is shown in
Fig.~\ref{fig:call-graph}.

\subsection{Prefetching techniques and strategies for high-latency
  environments}
Efficient data loading is essential for optimizing training
throughput, particularly in high-latency network environments. A
common approach leverages the fact that, even if the dataset is
shuffled at the start of each epoch, the permutation is predetermined
and all the future requests are known at the beginning of each
epoch. This allows to apply prefetching techniques, enabling
subsequent batches to be retrieved while the GPUs process the current
one, thereby minimizing idle time and improving overall efficiency.

Our data loader features a prefetching mechanism with a configurable
number of batch buffers, designed to mask latencies of varying
magnitudes. Despite this, during preliminary tests over real
high-latency internet connections, we observed significant
underperformance compared to internal tests with artificially induced
latency using tc-netem~\cite{netem}. Traffic analysis indicated
significant bandwidth variability due to multiple TCP connections
traversing different network routes, some of which experienced
congestion, resulting in a wide disparity between the best and
worst-performing connections. This directly impacts batch loading
times: since images are retrieved in parallel but assembled in order,
the system must wait for the slowest connection before proceeding,
which ultimately slows down the entire process.

To address this issue, we implemented an out-of-order prefetching
strategy.  Given that DL training is robust to uniformly random
permutations of the dataset, we can concurrently request multiple
batches (e.g., 8) and reassemble them based on the arrival time of the
contained images. This approach reduces the impact of slow connections
by prioritizing images that arrive first. For this strategy to
function correctly, it is essential that labels are retrieved together
with their corresponding features, a requirement met by our
architecture, which retrieves both features and annotations with a
single query.

Further testing over real high-latency internet connections revealed
another, second-order, issue: an aggressive filling of the prefetch
buffers (e.g., 8 buffers per GPU across 8 GPUs) can cause a burst of
requests that temporaly overwhelms the network capacity, leading to
unstable throughput during buffer filling. To mitigate this, we can
stagger the prefetch requests over time. For example, instead of
front-loading all prefetch requests, we can request an extra batch
every four regular ones: for every four batches consumed, five new
ones are requested until the buffer is full. This approach limits the
increase in transient throughput to only 25\% above the steady-state
level.

These optimizations collectively enhance throughput and stabilize data
loading in high-latency environments, significantly improving resource
utilization, as detailed in Sec.~\ref{sec:evaluation}.

\section{Software evaluation and results}
\label{sec:evaluation}

\subsection{Availability and usage}
Our data loader is free software released under the Apache-2.0 license
and is accessible on GitHub at the following URL:
\url{https://github.com/crs4/cassandra-dali-plugin/}. The repository
includes a Docker container that provides a pre-configured environment
with DALI, Cassandra DB, a sample dataset for experimentation, and
comprehensive instructions for conducting further tests.
The repository provides scripts to streamline the ingestion of
datasets into Cassandra, either serially or in parallel via Apache
Spark. It also includes examples of multi-GPU DL training in PyTorch,
using both plain PyTorch and PyTorch Lightning. Additionally, it
features tools for automatic dataset splitting and high-performance
inference using NVIDIA
Triton\footnote{\url{https://developer.nvidia.com/triton-inference-server}}. This
setup enables clients to request inference of images stored on a
remote Cassandra server to be processed on a different GPU-powered
remote server.

Listing~\ref{lst:file-reader} presents the Python code used to
initialize a typical DALI pipeline, which includes data loading via
the standard file reader, decoding, and standard preprocessing steps
such as resizing, cropping, and normalization. To use our custom data
loader, which supports data reading from a Cassandra-compatible DB,
one simply needs to replace the standard file reader with our module,
as demonstrated in Listing~\ref{lst:cassandra-reader}. The modified
lines are highlighted in blue.

\lstset{language=python,
  showstringspaces=false,
  morekeywords={readers},
  moredelim=**[is][\color{crs4blue}]{£}{£},
}
\begin{lstfloat*}
  \caption{Initializing DALI pipeline using DALI standard file reader}
  \label{lst:file-reader}
  \begin{lstlisting}
@pipeline_def(batch_size=128, num_threads=4, device_id=device_id)
def get_dali_pipeline():
    £images, labels = fn.readers.file(name="Reader",
      file_root="/data/imagenet/train")£
    labels = labels.gpu()
    images = fn.decoders.image(images,  device="mixed",
      output_type=types.RGB)
    images = fn.resize(images, resize_x=256, resize_y=256)
    images = fn.crop_mirror_normalize(images, dtype=types.FLOAT,
        output_layout="CHW", crop=(224, 224),
        mean=[0.485 * 255, 0.456 * 255, 0.406 * 255],
        std=[0.229 * 255, 0.224 * 255, 0.225 * 255],
    )
    return images, labels
  \end{lstlisting}
\end{lstfloat*}

\lstset{language=python,
  showstringspaces=false,
  morekeywords={cassandra,crs4},
  moredelim=**[is][\color{crs4blue}]{£}{£},
}
\begin{lstfloat*}
  \caption{Initializing DALI using our Cassandra-DALI plugin}
  \label{lst:cassandra-reader}
  \begin{lstlisting}
£uuids = list_manager.get_list_of_uuids(...)£

@pipeline_def(batch_size=128, num_threads=4, device_id=device_id)
def get_dali_pipeline():
    £images, labels = fn.crs4.cassandra(name="Reader",
        cassandra_ips=[1.2.3.4, 5.6.7.8],
        username="guest", password="test",
        table=imagenet.data_train, uuids=uuids,
        prefetch_buffers=16, io_threads=8
    )£
    labels = labels.gpu()
    images = fn.decoders.image(images,  device="mixed",
      output_type=types.RGB)
    images = fn.resize(images, resize_x=256, resize_y=256)
    images = fn.crop_mirror_normalize(images, dtype=types.FLOAT,
        output_layout="CHW", crop=(224, 224),
        mean=[0.485 * 255, 0.456 * 255, 0.406 * 255],
        std=[0.229 * 255, 0.224 * 255, 0.225 * 255],
    )
    return images, labels
  \end{lstlisting}
\end{lstfloat*}

\subsection{Comparative analysis at varying latencies}
\label{sec:comp-analys-syst}

We evaluated our network data loader alongside two state-of-the-art
competitors, leveraging Amazon EC2 instances located in different
geographical regions to introduce varying latencies. Specifically, we
tested data consumption in Oregon using an 8-GPU p4d.24xlarge node,
while storing images at locations characterized by the following
latencies:
\begin{itemize}
\item \textbf{Low:} Data stored in Oregon, round-trip time (RTT) < 1~ms.
\item \textbf{Medium:} Data stored in Northern California, RTT $\simeq$ 20~ms.
\item \textbf{High:} Data stored in Stockholm, Sweden, RTT $\simeq$ 150~ms.
\end{itemize}

While storing images on a different continent from the GPUs is not a
common practice, the high-latency scenario was included to highlight
the challenges posed by latency-induced bottlenecks. Such challenges
are expected to further intensify in the future, as computational power
and bandwidth improve, whereas latency remains constrained by physical
limits.

For our experiments, we utilized the standard ImageNet-1k dataset,
described in Table~\ref{tab:test-specs}. The dataset was prepared in
the formats required by each data loader and stored on a single node
equipped with four NVMe SSDs, configured as a single striped logical
volume for optimized data access. Specifically, we used r5dn.24xlarge
instances in Oregon and Stockholm, and the similar m6in.24xlarge
instance in Northern California, where the previous instance type was
not available. As the RAM capacity of these machines surpasses the
size of our test dataset (i.e., ImageNet-1k), we reserved memory to
maintain approximately only 70~GB of free RAM (i.e., half the size of
the dataset). This approach prevents dataset caching in main memory,
ensuring that data is consistently read from the disks during
testing. By doing so, we simulate conditions involving larger datasets
that exceed the available memory capacity.

\begin{table}[h]
  \caption{ImageNet-1k dataset properties and test parameters}
  \label{tab:test-specs}
  \begin{tabular}{lr@{~}l}
    \toprule
    \textbf{Training set} & 1,281,167 & images \\
    \midrule
    \textbf{Average image size} & 115 & kB \\
    \midrule
    \textbf{Total size} & 147 & GB  \\
    \midrule
    \textbf{Batch size} & 512 & images \\
    \bottomrule
  \end{tabular}
\end{table}

\begin{table}[h]
  \caption{Overview of the tested data loaders.}
  \label{tab:tested-data-loaders}
  \begin{tabular}{llll}
    \toprule
    \textbf{Data Loader} & \textbf{Version} & \textbf{Data Storage} & \textbf{Granularity} \\
    \midrule
    \textbf{Cassandra-DALI} & 1.2.0 & ScyllaDB (Cassandra-compatible) & Single
    image\\
    \midrule
    \textbf{MosaicML SD} & 0.10.0 & MinIO S3-compatible server & Record file (MDS)\\
    \midrule
    \textbf{tf.data service} & 2.16.1 & Filesystem on remote node & Record file (TFRecord) \\
    \bottomrule
  \end{tabular}
\end{table}

Three data loaders, summarized in Table~\ref{tab:tested-data-loaders},
were compared in this study: our Cassandra-DALI data loader (with data
stored in high-performance ScyllaDB), MosaicML SD (with data hosted on
an S3 MinIO server), and TensorFlow's tf.data service (with data
stored as TFRecords in the filesystem). All servers were hosted on the
same node within Docker containers, with all data residing on the same
logical volume.  The test code, including the Dockerfiles, is
available in the following GitHub repository, under the \emph{paper}
branch: \url{https://github.com/fversaci/cassandra-dali-plugin/}.

We conducted two experiments to evaluate performance under varying
latencies:
\begin{itemize}
\item \textbf{Tight-loop read:} This benchmark assesses raw
  data-loading capabilities by maximizing data reads without GPU
  processing or image decoding.
\item \textbf{Training:} A standard PyTorch multi-GPU ResNet-50
  training workload, including image decoding and preprocessing steps
  such as resizing, normalization, and cropping.
\end{itemize}

It is important to note that the tight-loop read test utilizes a
single data loader, whereas the training process employs a separate
data loader for each GPU. Consequently, the tight-loop read test
establishes an upper bound on the data throughput that can be consumed
by a single GPU.

To minimize AWS usage costs, each data loader was evaluated in a
single test run for up to four epochs. The experimental results
presented below compare the performance of the data loaders under
varying latency conditions.

\subsubsection{Tight-loop reading}
\label{sec:tight-loop-reading}

As a baseline for comparing network data loaders, we measured the
performance of NVIDIA DALI when reading images stored as TFRecords
from the local filesystem, without performing image decoding. This
configuration achieved a data throughput of 7.4~GB/s.

The p4d.24xlarge instance offers a total network bandwidth of
100~Gb/s; however, only half of this bandwidth is accessible through
its public
interface\footnote{\url{https://docs.aws.amazon.com/AWSEC2/latest/UserGuide/ec2-instance-network-bandwidth.html}}.
Thus, the maximum raw bandwidth available for data transfers in our
tests was limited to 50~Gb/s (6.25~GB/s).
As shown in Fig.~\ref{fig:loop-perf} and Tab.~\ref{tab:loop-perf}, our
data loader nearly saturated the available bandwidth when reading from
ScyllaDB in both local and medium-latency settings, achieving a
throughput of approximately 6~GB/s in each case. In the high-latency
setting, throughput decreased to around 4~GB/s.

In contrast, MosaicML SD demonstrated significantly lower performance,
with throughput measured at 326~MB/s in the low-latency setting,
308~MB/s in the medium-latency setting, and 203~MB/s in the
high-latency setting. tf.data service exhibited better performance
than MosaicML SD in the low-latency configuration, achieving a
throughput of 437~MB/s. However, its performance degraded
substantially in higher-latency environments, with throughput dropping
to 57~MB/s in the medium-latency setting and just 12~MB/s in the
high-latency setting.

\subsubsection{Training}
\label{sec:training}

For a data loader to be effective, its performance must integrate
smoothly into the DL pipeline, ensuring that tensors are efficiently
delivered to DL engines without introducing bottlenecks or delays. To
evaluate this, we assessed the performance of data loaders within a
standard image classification training workflow using the ResNet-50
architecture.
Due to the differences in training performance between TensorFlow and
PyTorch, we focused exclusively on one framework to ensure a fair
comparison. Given that MosaicML SD demonstrated superior performance
compared to tf.data service in medium- and high-latency settings in
the previous evaluation, we chose to test it against our data loader
in a training using PyTorch.

To establish a performance upper bound, we first performed training
using a fixed input tensor, thereby eliminating the overhead
associated with data loading and preprocessing. This setup enabled us
to measure the maximum achievable data throughput during
training. Specifically, we recorded the number of images processed per
second on a single GPU and across all 8 GPUs during multi-GPU
training. Our results indicate that a single NVIDIA A100 GPU consumes
about 1450 images/s, while an 8-GPU configuration reaches 11200
images/s. Given the average ImageNet-1k training image size of 115~kB,
the data loaders must sustain a steady throughput of approximately
1.3~GB/s to meet these requirements.

\begin{figure}
  \centering
  \subcaptionbox{Normalized tight-loop reading throughput.\label{fig:loop-perf}}  {
    \includegraphics[height=43mm]{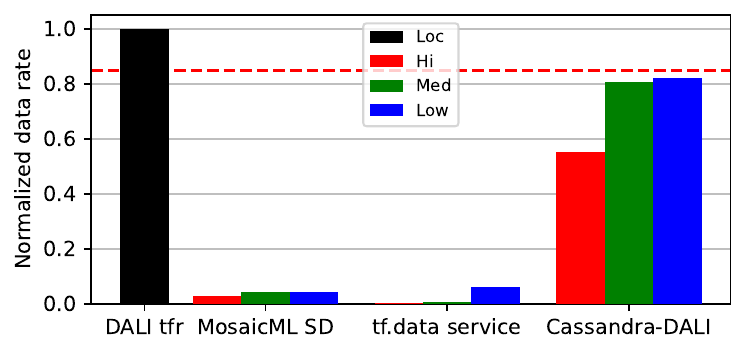}
  }\hfill
  \subcaptionbox{Normalized train reading throughput.\label{fig:train-perf}}
  {
    \includegraphics[height=43mm]{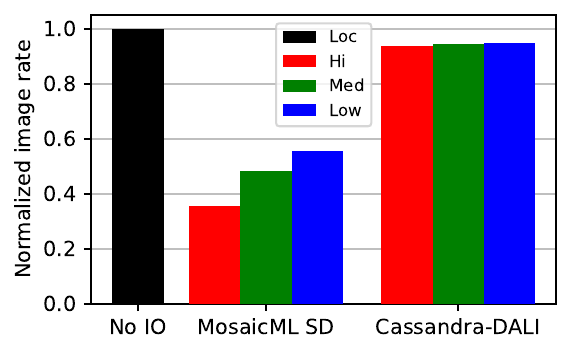}
  }
  \caption{Normalized reading throughput at varying latencies. The
    red, dashed line in (a) shows the available network bandwidth.}
\end{figure}

As demonstrated in Fig.~\ref{fig:train-perf} and
Tab.~\ref{tab:train-perf}, the MosaicML SD data loader is unable to
sustain the throughput required to fully utilize all 8 GPUs, achieving
57\%, 49\%, and 33\% of the target throughput under low, medium, and
high-latency conditions, respectively. In contrast, our data loader
achieves 94\%, 95\%, and 96\% of the theoretical upper bound in these
settings.

\begin{table}[h]
  \caption{Tight-loop reading at varying latencies. The table shows
    average epoch data throughput and standard deviation (omitted when
    data is insufficient). Epoch throughput is calculated as the
    dataset size divided by epoch duration.}
  \label{tab:loop-perf}
  \begin{tabular}{crrrr}
    \toprule
    \multirow{2}{*}{\textbf{Data loader}} & \multicolumn{4}{c}{\textbf{Throughput (MB/s)}}\\
    & \textbf{Local} & \textbf{Low-lat} & \textbf{Med-lat} & \textbf{Hi-lat}\\
    \midrule
    DALI TFRecord & 7381 $\pm$ 108 \\
    \midrule
    MosaicML SD & & 326 $\pm$ 14\phantom{1} & 308 $\pm$ 15\phantom{1} & 203 $\pm$ 8\phantom{12} \\
    \midrule
    tf.data service & & 437 $\pm$ 5\phantom{12} & 57 \phantom{$\pm$ 123}
      & 12 \phantom{$\pm$ 123} \\
    \midrule
    \textbf{Cassandra-DALI} & & 6066 $\pm$ 147 & 5957 $\pm$ 115 & 4081 $\pm$ 337 \\
    \bottomrule
  \end{tabular}
\end{table}

\begin{table}[h]
  \caption{Pytorch ResNet-50 training at varying latencies. The table
    shows average epoch image throughput and standard deviation
    (omitted when data is insufficient). Epoch throughput is
    calculated as the dataset size divided by epoch duration.}
  \label{tab:train-perf}
  \begin{tabular}{crrrr}
    \toprule
    \multirow{2}{*}{\textbf{Data loader}} & \multicolumn{4}{c}{\textbf{Throughput (img/s)}}\\
    & \textbf{Local} & \textbf{Low-lat} & \textbf{Med-lat} & \textbf{Hi-lat}\\
    \midrule
    No I/O & 11199 $\pm$ 312 \\
    \midrule
    MosaicML SD & & 6209 $\pm$ 89\phantom{1} & 5424 $\pm$ 735 & 3992 $\pm$ 5\phantom{12} \\
    \midrule
    \textbf{Cassandra-DALI} & & 10608 $\pm$ 113 & 10587 $\pm$ 300 & 10485 $\pm$ 98\phantom{1} \\
    \bottomrule
  \end{tabular}
\end{table}

\subsection{Impact of prefetching and database choice on data loader
  performance}
Finally, we conducted further tests to better investigate our data
loader's performance, focusing specifically on the impact of the
proposed out-of-order prefetching optimization. Additionally, we
analyzed how the choice of the underlying database -- either Cassandra
or ScyllaDB -- affects data loading performance.

\subsubsection{Impact of out-of-order, incremental prefetching}
We evaluated the tight-loop reading performance under high-latency
conditions, comparing results with and without our out-of-order,
incremental prefetching optimization.

High-latency, high-bandwidth internet communications are prone to
significant variability in TCP throughput. In fact these conditions
exacerbate the effects of packet loss, as TCP congestion control
mechanisms respond conservatively to retransmissions and recover
slowly due to extended RTTs, resulting in reduced throughput
\cite{tcp-congestion,tcp-cubic}.

\begin{figure}
  \centering
  \subcaptionbox{In-order prefetching.\label{fig:noooo-batch-time}}  {
    \includegraphics[width=.47\linewidth]{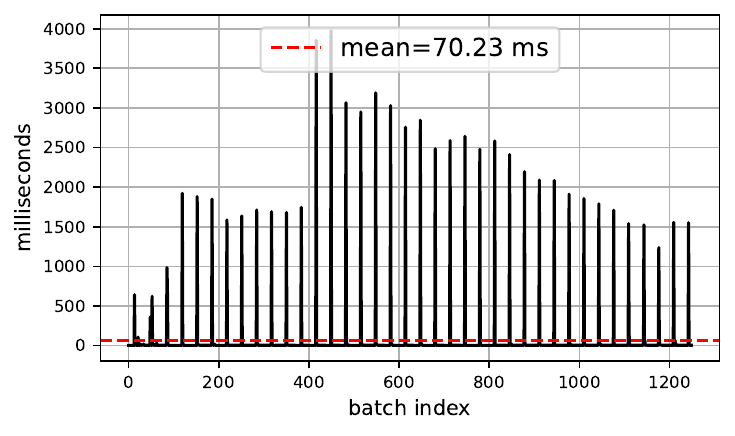}
  }\hfill
  \subcaptionbox{Out-of-order prefetching.\label{fig:ooo-batch-time}}
  {
    \includegraphics[width=.47\linewidth]{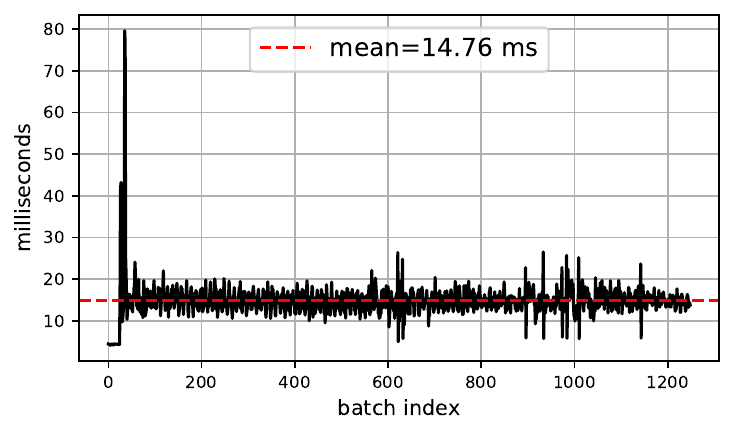}
  }
  \caption {Comparison of batch loading times in high-latency networks.}
  \label{fig:ooo-vs-noooo-batch-time}
\end{figure}

Figure~\ref{fig:ooo-vs-noooo-batch-time} highlights the significant variance in
batch loading times between in-order and out-of-order prefetching. In
the in-order case (Figure~\ref{fig:noooo-batch-time}),
when the prefetching queue is exhausted,
the system experiences delays of up to several seconds while waiting
for all transfers, including those over congested routes, to
complete. This results in a cyclical pattern: once all transfers are
completed, the queue is refilled, but it is quickly depleted again,
triggering a new cycle. In contrast, the out-of-order approach
(Figure~\ref{fig:ooo-batch-time}) maintains a highly consistent batch
loading time, staying always below 30~ms after the initial transient period.

\begin{figure}
  \centering
  \includegraphics[width=\linewidth]{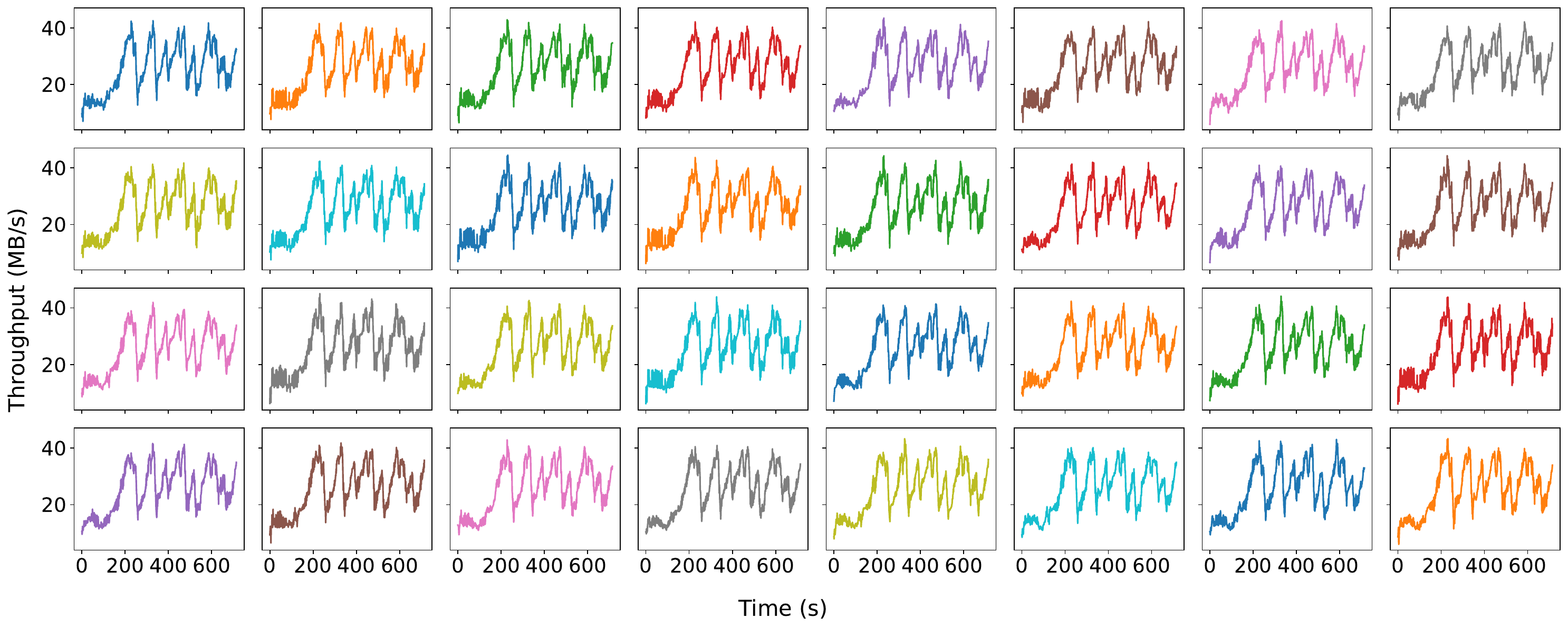}
  \caption{Transfer rate of 32 concurrent, high-latency TCP
    connections. In-order prefetching.}
  \label{fig:tcp-conn-no-ooo}
\end{figure}

Figure~\ref{fig:tcp-conn-no-ooo} illustrates the throughput over time
of each of the 32 TCP connections utilized by our data loader when
employing the standard in-order prefetching mechanism. The throughput
curves exhibit a strong correlation, highlighting that simultaneous
transfers are constrained by the in-order batch assembly process,
since the system must wait for the slowest transfer to finish before
dispatching a batch to the DL pipeline and requesting a new batch from
the database. As a result, the throughputs tend to converge and the
aggregated throughput exhibits considerable fluctuations, ranging
roughly from 300~MB/s to 1300~MB/s.

\begin{figure}
  \centering
  \includegraphics[width=\linewidth]{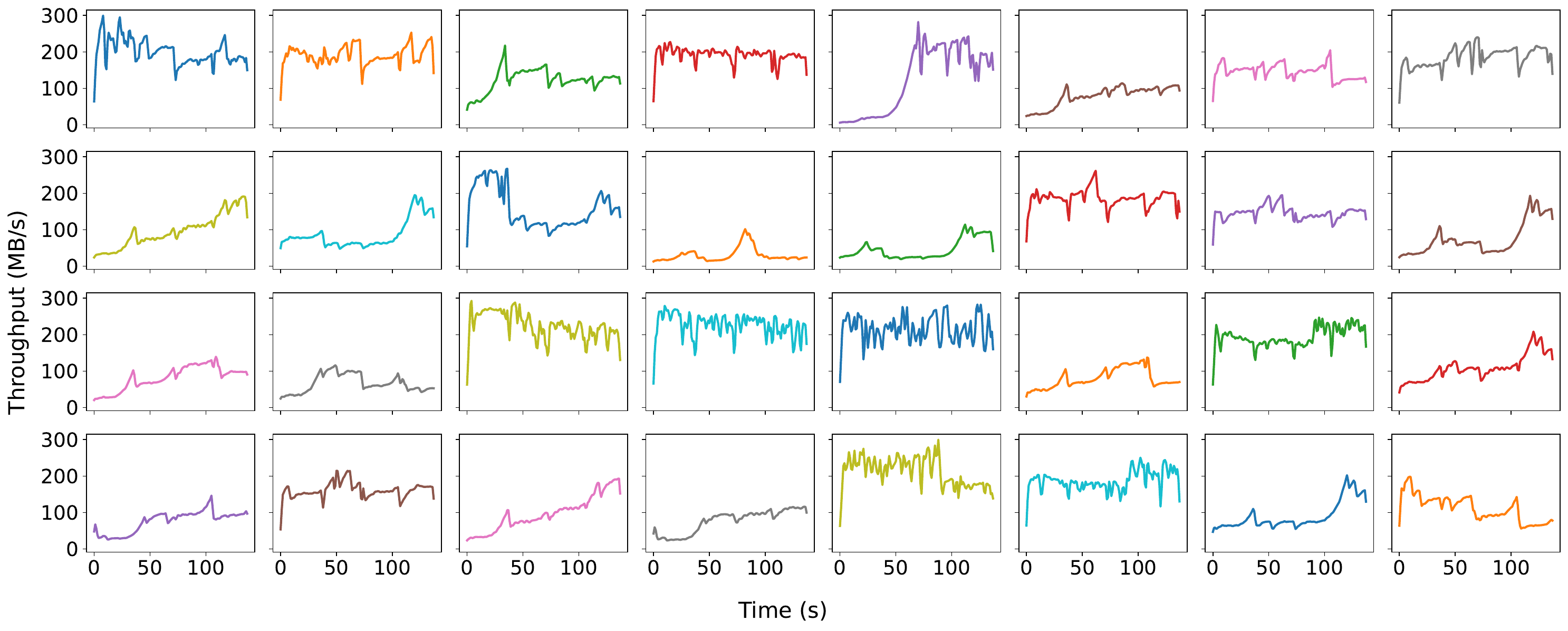}
  \caption{Transfer rate of 32 concurrent, high-latency TCP
    connections. Out-of-order prefetching.}
  \label{fig:tcp-conn-ooo}
\end{figure}

In contrast, relaxing this in-order constraint allows transfers to
proceed independently, as shown in Figure~\ref{fig:tcp-conn-ooo}. In
this approach, batches are formed as soon as a sufficient number of
images are available, irrespective of their originating
connections. This optimization significantly enhances overall
throughput, resulting in higher and more consistent performance,
maintaining an average throughput of approximately 4~GB/s.

\subsubsection{Cassandra vs ScyllaDB}
\begin{figure}
  \centering
  \includegraphics[width=.4\linewidth]{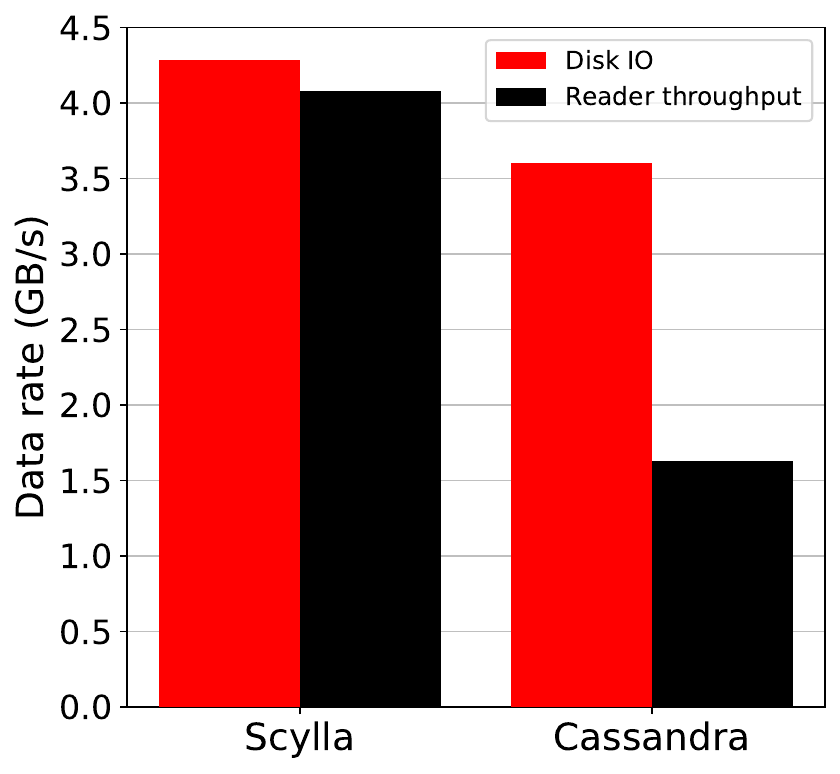}
  \caption{Comparison of disk and network throughputs for Cassandra
    and ScyllaDB (tight-loop reading test).}
  \label{fig:cass-scylla}
\end{figure}

The tight-loop reading test under high-latency conditions was also
performed using Cassandra as the storage backend for images, replacing
ScyllaDB. As illustrated in Fig.~\ref{fig:cass-scylla}, Cassandra
achieved a throughput of 1.6~GB/s, significantly lower than the
4.0~GB/s observed with ScyllaDB, highlighting the superior performance
of the latter. Notably, Cassandra exhibited a disk I/O rate
considerably higher than its achieved data throughput (3.6~GB/s versus
1.6~GB/s), likely attributable to differences in its block-reading
strategy compared to ScyllaDB.

\section{Conclusion}
The exponential growth in GPU computational power has enabled
unprecedented advancements in deep learning, particularly for
large-scale applications. However, the increasing discrepancy between
processing throughput and data access latencies has introduced
significant challenges in ensuring efficient data movement and storage
management. Our proposed solution addresses these challenges by
integrating scalable NoSQL databases with a high-performance data
loader optimized for image-based DL tasks.

The key contributions of our work include a novel data loader designed
to leverage state-of-the-art prefetching strategies, including
out-of-order prefetching mechanisms that mitigate the impact of
network variability and congestion, thereby maximizing data loading
efficiency. By coupling data with metadata and employing a
database-driven architecture, our implementation provides a scalable,
flexible, and consistent solution for managing DL datasets. Through
comprehensive evaluations under varying latency conditions, we
demonstrated the effectiveness of our approach, achieving significant
improvements in throughput and stability, particularly in high-latency
environments. Comparative analyses further validate the robustness of
our method against existing state-of-the-art data loading techniques.

The source code for our implementation is publicly available,
providing a resource for further research and practical deployment in
diverse DL scenarios.

\section*{Acknowledgment}

This work was supported by the Italian Ministry of Health under the
program grant H2ub (\emph{Hybrid Hub: Cellular and computation
models, micro- and nano-technologies for personalized innovative
therapies}, project code T4-AN-10) and by the Regione Autonoma
della Sardegna, Sardegna Ricerche, under the program grant XDATA.

\bibliography{netdl}

\end{document}